\documentclass[pra, aps, twocolumn, superscriptaddress, nofootinbib, tightenlines, nobibnotes, showpacs, 10pt]{revtex4-1}
\usepackage{natbib}
\usepackage{slashed}
\usepackage{graphicx}
\usepackage{subfigure}
\usepackage[usenames, dvipsnames]{color}
\usepackage{graphics}
\usepackage{hyperref}
\usepackage{bm}
\usepackage{amsmath}
\usepackage{xcolor}
\usepackage{amsfonts}

\definecolor{lapislazuli}{rgb}{0.15, 0.38, 0.61}
\definecolor{carmine}{rgb}{0.59, 0.0, 0.09}
\definecolor{carminered}{rgb}{1.0, 0.0, 0.22}

\usepackage{multirow}

\hypersetup{backref,
colorlinks=true,
linkcolor=lapislazuli,
linktoc=page,
citecolor=lapislazuli,
urlcolor=lapislazuli}

\everymath{\displaystyle}

\begin{document}

\title{Bound-state spectrum of an impurity in a quantum vortex}

\author{Jo\~ao E. H. Braz}

\affiliation{CeFEMA, Instituto Superior T\'ecnico, Universidade de Lisboa, Lisboa,
Portugal}

\author{H. Ter\c{c}as}

\affiliation{IPFN, Instituto Superior Técnico, Universidade de Lisboa, Lisboa,
Portugal}
\begin{abstract}
We consider the problem of finding the bound-state spectrum of an
impurity immersed in a weakly interacting two-dimensional Bose-Einstein
condensate supporting a single vortex. We obtain approximate expressions
for the energy levels and show that, due to the finite size of the
condensate, the impurity can access only a finite number of physical
bound states. By virtue of the topological quantization of the vorticity
and of the emergence of the Tkachenko lattice, this system is promising
as a robust and scalable platform for the realization of qubits. Moreover,
it provides a potentially new paradigm for polaron physics in Bose-Einstein
condensates and a glimpse towards the study of quantum turbulence
in low-dimensionality systems. 
\end{abstract}
\maketitle

\section{Introduction}

The study of quantum many-body systems has a history of unveiling
remarkable new physics upon the inclusion of \emph{impurities} - particles
distinct from those comprising the majority, due to their mass, spin,
or charge. The understanding of such composite systems~\cite{gordon_2000,Devreese_2014,Dykman_2015,vojta_2019},
along with the development of appropriate theoretical techniques,
has been not only an enlightening process but also a necessary one,
since the existence of impurities is inevitable in realizations of
any physical system that condensed matter theory may aim to describe~\cite{devreese_2003,drescher_2019,yoshida_2018}.

A paradigmatic example of the presence of impurities is the polaron,
a quasi-particle resulting from the hybridization between an electron
and a lattice phonon \cite{frohlich_1950,grimvall_1981,stamatis_1995}.
The weak electron-phonon coupling, the so-called Fhr\"olich polaron,
initiated the understanding of phonon-mediated superconductivity \cite{swartz_1475,appel_1969,gerlach_1991,kudinov_2002},
while recent progress in analytical and numerical techniques has allowed
the description of more generic regimes \cite{grimaldi_2008,grimaldi_2010,frank_2014}.

Though firmly rooted in the phenomenology of solid-state physics,
interest in analogue models of polaron physics by immersion of impurities
in Bose-Einstein condensates (BEC) has grown in recent years~\cite{tempere_2009,catani_2012,scelle_2013,PhysRevLett.117.055302,drescher_2019,mistakidis_2019,ardila_2019},
a fact to be partially attributed to the high degree of controllability
in ultracold-atom experiments. In one hand, impurities are ubiquitous
in superfluid liquid Helium experiments and known to be at the origin
of the pinning of vortex lines \cite{gordon_2000,pelmenev_2016}.
This effect has been crucial for the experimental observation of vortices
in superfluids by means of spectroscopic techniques, and therefore
to the study of quantum turbulence \cite{barenghi_2014,PhysRevB.99.140505,Pshenichnyuk:2017aa};
in BECs, on the other hand, vortices can be produced in a controlled
fashion \cite{Abo_Shaeer_2001,Lin_2009,Madison:2000aa,Matthews:1999aa,PhysRevLett.97.170406,Schweikhard_2004},
ranging from single-vortex realizations \cite{PhysRevLett.97.170406,Matthews:1999aa}
to the production of Tkachenko lattices \cite{Abo_Shaeer_2001,Schweikhard_2004}.
Most importantly, their stability is linked to a topological invariant
quantizing the fluid angular momentum, making them as long-lived as
the condensate itself \cite{svidzinsky_2000}. In quasi-one dimensional
BECs, the interaction of impurities with dark solitons has shown to
be sufficiently rich to make possible a variety of applications in
quantum information theory \cite{Shaukat:2017aa,Shaukat:2018aa,Shaukat:2019aa}.

In this work, we investigate the eigenvalue problem of an impurity
bounded to a single vortex in a quasi two-dimensional (2D) condensate.
In particular, we show that there is a tunable, finite number of bound
states, making it a promising scheme for the realization of a qubit.
This comes with advantages in respect to the one dimensional ``dark-soliton
qubit\char`\"{} considered in \cite{Shaukat:2017aa}, as vortices
are more stable in respect to the excitations of sound waves and offer
additional flexibility regarding scalability, since the number of
vortices that can be produced in a 2D dimensional rotating BEC (state-of-the-art
Tkachenko lattices contain up to hundred vortices) is much larger
than the few tens of solitons that one can produce in 1D traps.

This paper is organized as follows: In Sec.~\ref{sec:Stationary-solutions-of},
we set the basic equations describing the stationary vortex-impurity
problem, where the vortex acts as a potential trapping the impurity.
Upon establishing a variational approximation for the vortex profile,
we solve the eigenvalue problem in Sec.~\ref{sec:Bound-state-spectrum}.
In Sec.~\ref{sec:Finite-size-effects-and}, we show how the finite
size of the condensate determines which bound states are physical.
In Sec.~\ref{sec:Experimetnal-considerations},
we discuss some experimental considerations towards the realization
of this system. Finally, in Sec.~\ref{sec:Conclusion}, a discussion
of the physical results and some concluding remarks are enclosed.

\section{Stationary solutions of the weakly interacting vortex-impurity system\label{sec:Stationary-solutions-of}}


\subsection{The eigenvalue problem}

We begin by considering the Gross-Pitaevskii equations (GPE) that
describe the quasi two-dimensional (2D) BEC coupled to a single impurity,
as 
\begin{align}
\text{i}\hbar\partial_{t}\psi_{1} & =-\frac{\hbar^{2}}{2m_{1}}\nabla^{2}\psi_{1}+g_{11}\left|\psi_{1}\right|^{2}\psi_{1}+g_{12}\left|\psi_{2}\right|^{2}\psi_{1}\label{eq:GPE-BEC}\\
\text{i}\hbar\partial_{t}\psi_{2} & =-\frac{\hbar^{2}}{2m_{2}}\nabla^{2}\psi_{\text{2}}+g_{12}\left|\psi_{1}\right|^{2}\psi_{2}\,.\label{eq:GPE-imp}
\end{align}
Here, $m_{1}$ and $m_{2}$ respectively denote the BEC particle and
the impurity masses, $g_{11}$ is the interaction strength of the
BEC particles and $g_{12}=g_{21}$ stems from the BEC-impurity interaction.
For definiteness, we consider the case of repulsive interactions only,
$g_{11}>0$ and $g_{12}>0$. In the quasi two-dimensional situation,
$\xi=\hbar/\sqrt{2n_{0}m_{1}g_{11}}$ is the healing length, defining
the typical size of the vortex core, where $n_{0}=N/A$ is the surface
density of the condensate, with $N$ being the total number of BEC
particles and $A$ the total area. Unless stated contrary, we perform
the following calculations in the thermodynamic limit, $N\rightarrow\infty$
and $A\rightarrow\infty$, while taking $n_{0}$ constant. Within
the present considerations, we obtain the stationary eigenvalue problem
by extracting the time-dependence of Eqs. \eqref{eq:GPE-imp} as 
\begin{align}
\psi_{1}(\mathbf{r},t) & =\sqrt{n_{0}}\exp(-\frac{\text{i}}{\hbar}\mu t)\Phi(\mathbf{r})\,,\\
\psi_{2}(\mathbf{r},t) & =\frac{1}{\sqrt{\xi^{2}}}\exp(-\frac{\text{i}}{\hbar}Et)\Psi(\mathbf{r}),
\end{align}
where $\mu$ is the BEC chemical potential and $E$ some parameter
of the impurity, to be defined later on. At zeroth order, i.e. by
neglecting the effect of the impurity on the BEC dynamics, we defined
the chemical potential as 
\begin{equation}
\mu\simeq\frac{\hbar^{2}}{2m_{1}\xi^{2}}=n_{0}g_{11}\,.
\end{equation}
As such, by performing the substitution $\mathbf{r}\rightarrow\mathbf{r}/\xi$,
Eqs. \eqref{eq:GPE-BEC} can be conveniently recast in a dimensionless
form 
\begin{align}
\nabla^{2}\Phi+\left(1-\left|\Phi\right|^{2}-\kappa^{2}\left|\Psi\right|^{2}\right)\Phi & =0\label{eq:red-BEC}\\
\nabla^{2}\Psi+\left(\epsilon-\gamma^{2}\left|\Phi\right|^{2}\right)\Psi & =0,\label{eq:red-imp}
\end{align}
with $\epsilon=2m_{2}\xi^{2}E/\hbar^{2}$, and the two parameters
controlling the relative strength of the intra- and inter-particle
interactions 
\begin{equation}
\gamma^{2}=\frac{g_{12}m_{2}}{g_{11}m_{1}}\,,\quad\kappa^{2}=\frac{2m_{1}g_{12}}{\hbar^{2}}.\label{eq:essential-params}
\end{equation}
As we can see from Eqs.~\eqref{eq:red-BEC} and~\eqref{eq:red-imp},
there is a limit in which we can neglect the effect of the impurity
on the BEC: making 
\begin{equation}
\frac{\kappa^{2}}{\gamma^{2}}=\frac{m_{1}}{m_{2}}\frac{1}{n_{0}\xi^{2}}\rightarrow0\,,\label{eq:decoup-condition}
\end{equation}
in the sense that $\kappa^{2}$ vanishes as $\gamma^{2}$ remains
finite, Eq.~\eqref{eq:red-BEC} becomes decoupled from $\Psi$. In
that limit, we may consider that the density $\vert\Phi\vert^{2}$
acts as a potential of depth $\gamma^{2}$ for $\Psi$ in Eq.~\eqref{eq:red-imp}.
In this weakly-interacting regime, we may handle the problem for the
impurity as a linear one, in the sense that Eq.~\eqref{eq:red-imp}
amounts to the time-independent Schr\"odinger equation 
\begin{equation}
H_{\text{imp}}\Psi=\epsilon\Psi\,,\label{eq:Schr-prob}
\end{equation}
with $H_{\text{imp}}$ being the effective Hamiltonian, 
\begin{equation}
H_{\text{imp}}=-\nabla^{2}+\gamma^{2}\left|\Phi_{0}\right|^{2}\,,
\end{equation}
where $\Phi_{0}$ is the solution of Eq. \eqref{eq:red-BEC} for $\kappa^{2}=0$.


\subsection{Vortex solution and the variational approximation}

\begin{figure}
\includegraphics{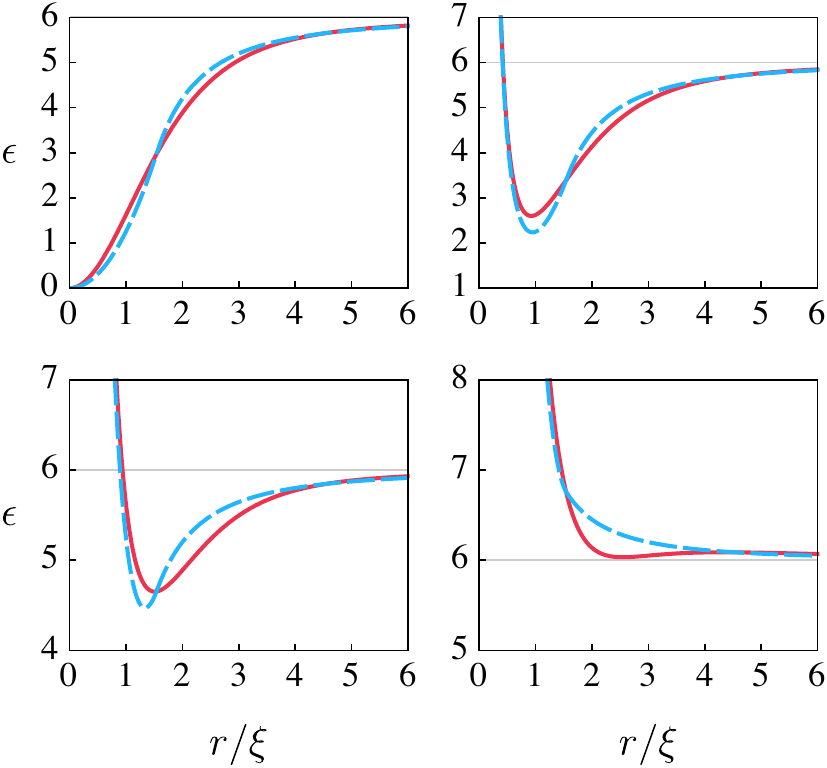} \caption{(color online) Comparison of effective one-dimensional radial potentials
for the exact solution (solid red) and the variational approximation
of Eq.~\eqref{eq:piecewise-var} (dashed blue), for $\ell=0$ (upper-left),
$\ell=1$ (upper-right), $\ell=2$ (lower-left) and $\ell=3$ (lower-right),
for a value $\gamma^{2}=6$ (horizontal grey line) of the potential
depth. We see that for $\ell=3$ the effective potentials are entirely
above this line, illustrating the relation between the potential depth
and angular momentum, and suggesting a loss of bound states for $\gamma^{2}$
decreasing below some threshold. \label{fig:Comparison-of-effective}}
\end{figure}

We are firstly interested in obtaining solutions to Eq. \eqref{eq:red-BEC}
in the limit $\kappa\rightarrow0$, 
\begin{equation}
\nabla^{2}\Phi_{0}+\left(1-\left|\Phi_{0}\right|^{2}\right)\Phi_{0}=0\,,\label{eq:rigid-vortex}
\end{equation}
which, in two dimensions, contains vortex solutions of the BEC \cite{pitaevskii_2003,pethick_smith_2008}.
Here, we particularize to circularly symmetric solutions, i.e. configurations
comprised of a single vortex at the origin. In polar coordinates,
they read 
\begin{equation}
\Phi_{0}(r,\varphi)=\phi(r)e^{in\varphi},\quad\phi(r)\equiv\left|\Phi_{0}(r,0)\right|,
\end{equation}
with $n$ being an integer, also known as the vortex charge. With
this prescription, Eq~\eqref{eq:rigid-vortex} then becomes 
\begin{equation}
\frac{\text{d}^{2}\phi}{\text{d}r^{2}}+\frac{1}{r}\frac{\text{d}\phi}{\text{d}r}-\frac{n^{2}}{r^{2}}\phi+\left(1-\phi^{2}\right)\phi=0\,.\label{eq:circ-vortex-rad}
\end{equation}
The solution for $n=0$ is the trivial one (i.e. the homogeneous condensate),
whereas solutions with $\left|n\right|>1$ are energetically unstable
- multiply charged vortices decay in singly charged ones \cite{PhysRevA.70.043610,PhysRevA.96.023617}.
In the following, we consider the case $\left|n\right|=1$; though
this is the simplest non-trivial solution of~\eqref{eq:rigid-vortex},
there are no known closed-form solutions of Eq.~\eqref{eq:circ-vortex-rad}~\cite{Manton:2004aa}.
Instead, we consider the asymptotic behaviour as $r\rightarrow0$
and $r\rightarrow+\infty$~\cite{Manton:2004aa} 
\begin{align}
r\rightarrow0: & \quad\phi\propto r+\mathcal{O}(r^{3})\,,\label{eq:asymp-r-to-0}\\
r\rightarrow+\infty: & \quad\phi=1-\frac{1}{2r^{2}}-\mathcal{O}(r^{-4})\,.\label{eq:asymp-r-to-inf}
\end{align}
These asymptotics provides us with criteria to look for variational
approximations to the solution of Eq. \eqref{eq:circ-vortex-rad}
\cite{Konishi:2009aa}. In the present work, we consider a one-parameter family of trial functions
of the form 
\begin{equation}
\phi_{\alpha}(r)=\begin{cases}
\frac{1}{2}\alpha r & \,,\quad0<r\leq r_{\alpha}\\
\sqrt{1-\frac{1}{\alpha^{2}r^{2}}} & \,,\quad r>r_{\alpha}
\end{cases}\,,\label{eq:piecewise-var}
\end{equation}
where $r_{\alpha}=\alpha^{-1}\sqrt{2}$ and $\alpha$ is the variational
parameter; this function is constructed to be continuous to first
derivative for any $\alpha\in\mathbb{R}^{+}$. In turn, we find the
value $\alpha=\sqrt{5/6}\simeq0.913$ to provide the optimal variational
approximation within the family of functions of~\eqref{eq:piecewise-var}.
A comparison with the numerical solution of Eq.~\eqref{eq:circ-vortex-rad}
for $\left|n\right|=1$ is shown in Fig.~\ref{fig:Comparison-of-effective}.

\subsection{Impurity wave functions }

\label{subsec:Impurity-wave-functions}

We now substitute $\left|\Phi_{0}\right|$ for $\phi_{\alpha}$ in
Eq. \eqref{eq:Schr-prob}, in order to obtain \emph{approximate} solutions
for the impurity wave functions, and look for bound states, which
amounts to requiring $0<\epsilon<\gamma^{2}$. In polar coordinates,
this is done by separating 
\begin{equation}
\Psi(r,\varphi)=R(r)e^{\pm\text{i}\ell\varphi},\label{eq:var-separation-imp}
\end{equation}
with $\ell=0,1,2,\text{\ensuremath{\dots}}$ denoting the angular momentum. As for the radial part, it is solved in the two regions $\mathcal{R}_{1}=[0,r_{\alpha}]$ and $\mathcal{R}_{2}=(r_{\alpha},\infty)$. In the first region, we have 
\begin{equation}
-\frac{\text{d}^{2}R_{1}}{\text{d}r^{2}}-\frac{1}{r}\frac{\text{d}R_{1}}{\text{d}r}+\left(\frac{\ell^{2}}{r^{2}}+\frac{\alpha^{2}\gamma^{2}}{4}r^{2}\right)R_{1}=\epsilon R_{1}.\label{eq:region1-eq}
\end{equation}
Due to the term $\ell^{2}/r^{2}$, these solutions must scale as $R_{1}(r)\propto r^{\ell}$
as $r\rightarrow0$ for each value of $\ell$, and so the correct
solution is given by 
\begin{align}
R_{1}(r) & =\mathcal{A}\,r^{\ell}e^{-\frac{\alpha\gamma}{4}r^{2}}\times\nonumber \\
 & \times M\left(-\frac{\epsilon}{2\alpha\gamma}+\frac{\ell+1}{2},\ell+1;\frac{\alpha\gamma}{2}r^{2}\right),
\label{eq:region1-sol}
\end{align}
where $M(a,b;z)$ is the confluent hypergeometric function (CHF) with
parameters $a$ and $b$, also known as the first Kummer function~\cite{Olver:2010:NHM:1830479},
and $\mathcal{A}$ is a normalization constant. Some relevant properties
of this function are discussed in Appendix \ref{sec:Leading-order-of}.
In turn, for the second region, we get 
\begin{equation}
\frac{\text{d}^{2}R_{2}}{\text{d}r^{2}}+\frac{1}{r}\frac{\text{d}R_{2}}{\text{d}r}+\frac{\gamma_{\alpha}^{2}-\ell^{2}}{r^{2}}R_{2}=\left(\gamma^{2}-\epsilon\right)R_{2}\,,\label{eq:region2-eq}
\end{equation}
where $\gamma_{\alpha}=\gamma/\alpha$. Here, the regular solution
as required by the vanishing of the wave function as $r\rightarrow+\infty$ is given by 
\begin{equation}
R_{2}(r)=\mathcal{A}\,K_{\text{i}\lambda_{\ell}}(qr)\,,\label{eq:region2-sol}
\end{equation}
where $\lambda_{\ell}=\sqrt{\gamma_{\alpha}^{2}-\ell^{2}}$ and $q=\sqrt{\gamma^{2}-\epsilon}$
is a real number (for bound states), and $K_{\text{i}\lambda_{\ell}}$
is the modified Bessel function of the second kind of \emph{imaginary}
order $\lambda_{\ell}$~\cite{Olver:2010:NHM:1830479,Dunster:1990aa}. Note that the character of the order (i.e. real or imaginary) does not affect the regularity of the solution~\eqref{eq:region2-sol}
by itself~\cite{Olver:2010:NHM:1830479}, but it rather has a critical
role in the bound-state spectrum of the impurity.

\section{Bound-state spectrum\label{sec:Bound-state-spectrum}}

\begin{figure}[t]
\includegraphics{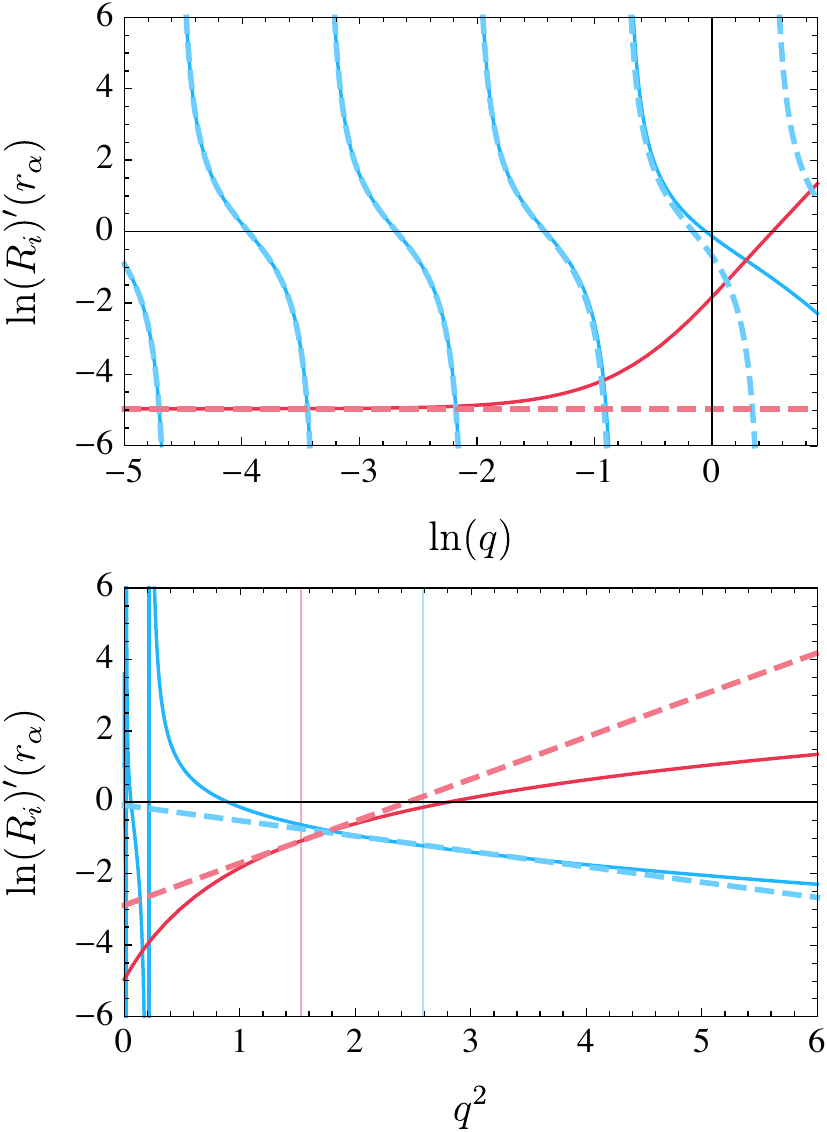}\caption{ (color online) An instance of Eq.~\eqref{eq:log-deriv} for $\ell=1$
and $\gamma^{2}=6$; the function corresponding to region 1 (2), the
deeper (shallower) end of the potential, is in thick red (blue); asymptotics
and approximations are the dashed lines of respective color. (Upper
panel) the choice of axis shows that there are infinitely many solutions
(given by intersections) as $q$ goes to zero, corresponding to shallow
states, and the contribution from $\mathcal{R}_{1}$ can be considered
constant, while the contribution from $\mathcal{R}_{2}$ is the RHS
of~\eqref{eq:log-deriv-rhs}; (lower panel) solutions for deep states
can be obtained by linearization and asymptotics of the logarithmic
derivatives, with the vertical lines indicating the linearization
points.\label{fig:An-instance-of}}
\end{figure}

The impurity spectrum in the vortex is obtained by requiring the continuity
of the logarithmic derivative at $r=r_{\alpha}$~\cite{Konishi:2009aa},
\begin{equation}
\left(\ln R_{1}\right)'(r_{\alpha})=\left(\ln R_{2}\right)'(r_{\alpha})\,,\label{eq:log-deriv}
\end{equation}
and solving for $q$. Fig.~\ref{fig:An-instance-of} shows plots
of these two functions. We find that Eq. \eqref{eq:log-deriv} has
no solutions for $\gamma_{\alpha}\leq\ell$, as shown in Appendix
\ref{sec:Inexistence-of-bound}, so we focus exclusively on the case
$\gamma_{\alpha}>\ell$. Consequently, the condition $\lambda_{\ell}=0$
yields critical values $\gamma_{\text{c},\ell}=\alpha\ell$ for the
onset of bound states of each angular momenta. Physically, we may
interpret this behavior as the impurity seeing the vortex potential
with an effective depth proportional to $\lambda_{\ell}^{2}$; indeed,
it is straightforward to check that, for each $\ell$, the effective
radial potential has the maximum depth $\alpha\gamma(\gamma_{\alpha}-\ell)$.
This suggests we can take $\lambda_{\ell}$ (or, further, the related
quantity $\Delta_{\ell}\equiv\gamma_{\alpha}-\ell$), as the relevant
scale of the problem. This point is illustrated in Fig. \ref{fig:Comparison-of-effective}.

Since we are interested in a vortex with few bound states, as it is
the case of a qubit, we focus on a parameter region 
\begin{equation}
0<\gamma_{\alpha}<3\,.\label{eq:depth-range-1}
\end{equation}
Further, in order to obtain qualitative results from Eq. \eqref{eq:log-deriv},
we posit that in the present parameter range the states can be distinguished
as shallow states and deep states: the former (latter) sit at the
edge (bottom) of the potential, thus being characterized by the
long-range, centripetal-like (short-range, harmonic oscillator-like)
profile of the vortex density, as implied by Eq.~\eqref{eq:asymp-r-to-inf}
(Eq.~\eqref{eq:asymp-r-to-0}) and approximated, respectively, by
each branch of Eq.~\eqref{eq:piecewise-var}.

\subsection{Shallow states}

In this case, we can expect $q=\sqrt{\gamma^{2}-\epsilon}\ll1$, prompting
us to look for solutions of Eq.~\eqref{eq:log-deriv} to leading
order in $q$. We thus use Eq. \eqref{eq:small-arg-K} to obtain 
\begin{equation}
\left(\ln R_{2}\right)'(r_{\alpha})=\frac{\lambda_{\ell}}{r_{\alpha}}\cot\left(\lambda_{\ell}\ln\frac{qr_{\alpha}}{2}-\varphi_{\ell}\right)+\mathcal{O}(q^{2}),\label{eq:log-deriv-rhs}
\end{equation}
where $\varphi_{\ell}\equiv\varphi(\lambda_{\ell})$ is related to
the gamma function by\textcolor{red}{{} }$\varphi(\lambda_{\ell})=\arg\Gamma(1+\text{i}\lambda_{\ell})$~\cite{Olver:2010:NHM:1830479,Dunster:1990aa}
(see~\ref{subsec:Bessel--of-imaginary} for further details). For
the LHS of Eq. \eqref{eq:log-deriv}, we have 
\begin{equation}
\left(\ln R_{1}\right)'(r_{\alpha})=-\frac{\lambda_{\ell}}{r_{\alpha}}\cot\theta_{\ell}+\mathcal{O}(q^{2})\,,\label{eq:2nd-line-lead-ord-lhs}
\end{equation}
with 
\begin{equation}
\theta_{\ell}=\tan^{-1}\left\{ \lambda_{\ell}\left[\Delta_{\ell}-2\gamma_{\alpha}\frac{M'\left(-\frac{\Delta_{\ell}}{2}+\frac{1}{2},\ell+1;\gamma_{\alpha}\right)}{M\left(-\frac{\Delta_{\ell}}{2}+\frac{1}{2},\ell+1;\gamma_{\alpha}\right)}\right]^{-1}\right\} \,,\label{eq:CapTheta}
\end{equation}
where the prime represents differentiation with respect to the argument.
The solution of Eq. \eqref{eq:log-deriv}, to leading order in $q$,
finally yields the bound-state spectrum for shallow states 
\begin{equation}
\frac{E_{p,\ell}}{n_{0}g_{12}}=1-\Lambda_{\ell}^{2}\exp\left(-\frac{2\pi p}{\lambda_{\ell}}\right)\,,\label{eq:bound-st-spect-ell>0}
\end{equation}
with $p$ the radial quantum number, taking on non-negative integer
values, and 
\begin{equation}
\Lambda_{\ell}=\frac{\sqrt{2}}{\gamma_{\alpha}}\exp\left(-\frac{(1-\delta_{\ell,0})\pi+\theta_{\ell}-\varphi_{\ell}}{\lambda_{\ell}}\right)\,,\label{eq:CapLambda_ell}
\end{equation}
where $\delta_{\ell,0}$ is a Kronecker delta. The description of
$p$ must take into account the number of deep states: the integers
$p$ indicate the the non-positive branches of the cotangent in~\eqref{eq:log-deriv-rhs}
at which~\eqref{eq:log-deriv} has solutions, as illustrated in Fig.~\ref{fig:An-instance-of};
as $\gamma$ increases, Eq.~\eqref{eq:log-deriv-rhs} no longer holds
for some values of $p$, as solutions transition from shallow to deep
states; hence, the counting of $p$ in~\eqref{eq:bound-st-spect-ell>0}
start from the number of deep states of angular momentum $\ell$, denoted here by $N_{\text{d}}^{(\ell)}$ and discussed below.

\subsection{Deep states}

Deep, or confined, states are considered according to the condition
\begin{equation}
\Delta_{\ell}>1,\label{eq:onset-deep-states}
\end{equation}
since, in this regime, it becomes possible for the first parameter
of the CHF in Eq.~\eqref{eq:region1-sol} to take on non-positive
values, 
\begin{equation}
-\frac{\epsilon}{2\alpha\gamma}+\frac{\ell+1}{2}\leq0\,,\label{eq:condition}
\end{equation}
from which~\eqref{eq:onset-deep-states} is obtained when $\epsilon=\gamma^{2}$,
the highest possible value for bound states. If Eq. \eqref{eq:condition}
is satisfied, the solutions to Eq. \eqref{eq:region1-sol} become
integrable on the whole plane, becoming the eigenfunctions of the
isotropic quantum harmonic oscillator (check Appendix \ref{subsec:Zeros-and-monotonicity}
for a more complete discussion around this issue). This observation
leads us to define the quantity 
\begin{equation}
\delta\epsilon_{1}\equiv-\frac{\epsilon}{2\alpha\gamma}+\frac{\ell+1}{2}+p\,,\label{eq:expand-point}
\end{equation}
for $p=0,1,2,\text{\ensuremath{\dots}}$, and to look for solutions
of Eq. \eqref{eq:log-deriv} to first order in $\delta\epsilon_{1}$
as the leading order contributions to its LHS. We thus require the first derivative of the CHF with respect to its
first parameter evaluated at non-positive integers; one approach to
this problem is presented in Appendix \ref{subsec:Derivative-with-respect}.

Furthermore, the criterion in Eq. \eqref{eq:onset-deep-states} can
be easily generalized to a more precise condition regarding the \emph{number}
of deep states, and we find that 
\begin{equation}
-1<\Delta_{\ell}-2N_{\text{d}}^{(\ell)}<1\,,\label{eq:num-dpst-criterion}
\end{equation}
for $N_{\text{d}}^{(\ell)}=0,1,2,\dots$, if there are $N_{\text{d}}^{(\ell)}$
deep states of angular momentum $\pm\ell$, accounting for the sign-degeneracy
in Eq. \eqref{eq:var-separation-imp}. Hence, within the parameter
range given in \eqref{eq:depth-range-1}, we have at most $N_{\text{d}}^{(0)}=N_{\text{d}}^{(1)}=1$
and $N_{\text{d}}^{(2)}=0$, corresponding to the state $p=0$, for
$\ell=0$ and for $\ell=1$, and none for $\ell=2$. According to
Eqs.~\eqref{eq:a-expand-chf} and~\eqref{eq:a0-1deriv-chf-int},
for $p=0$ the LHS of Eq.~\eqref{eq:log-deriv} becomes 
\begin{align}
 & \left(\ln R_{1}\right)'(r_{\alpha})=\nonumber \\
 & =-\frac{\Delta_{\ell}}{r_{\alpha}}+\frac{1}{r_{\alpha}}\frac{(\ell,\gamma_{\alpha})!}{\gamma_{\alpha}^{\ell}e^{-\gamma_{\alpha}}}\left(\ell+1-\frac{\epsilon}{\alpha\gamma}\right)+\mathcal{O}(\delta\epsilon_{1}^{2})\,,\label{eq:LHS-deep-st}
\end{align}
where $(\ell,\gamma_{\alpha})!$ is given in Eq.~\eqref{eq:inc-gamma-integer}.
The RHS of Eq.~\eqref{eq:log-deriv}, on the other hand, has a weak
dependence on $\epsilon$ for sufficiently large $\gamma$, since
at that point the solution~\eqref{eq:region2-sol} transitions from
an oscillatory to an exponentially decaying behavior, as discussed
in~\ref{subsec:Bessel--of-imaginary-1}. We can write $q=\lambda_{\ell}/r_{\alpha}+\delta\epsilon_{2}+\mathcal{O}(\delta\epsilon_{2}^{2})$,
and expand the RHS of Eq. \eqref{eq:log-deriv} to first order in
\begin{equation}
\delta\epsilon_{2}\equiv-\frac{\epsilon-\left(\gamma^{2}-\lambda_{\ell}^{2}/r_{\alpha}^{2}\right)}{2\lambda_{\ell}/r_{\alpha}}=\frac{1}{r_{\alpha}}\left(\frac{\gamma_{\alpha}^{2}+\ell^{2}}{2\lambda_{\ell}}-\frac{\epsilon}{\alpha^{2}\lambda_{\ell}}\right)\,.\label{eq:expand-point-2}
\end{equation}
We then find 
\begin{align}
 & \left(\ln R_{2}\right)'(r_{\alpha})=\nonumber \\
 & =-\frac{\lambda_{\ell}\mathcal{K}_{\ell}}{r_{\alpha}}-\frac{\lambda_{\ell}\mathcal{K}_{\ell}^{2}}{r_{\alpha}}\left(\frac{\gamma_{\alpha}^{2}+\ell^{2}}{2\lambda_{\ell}}-\frac{\epsilon}{\alpha^{2}\lambda_{\ell}}\right)+\mathcal{O}(\delta\epsilon_{2}^{2})\,,\label{eq:RHS-deep-st}
\end{align}
where 
\[
\mathcal{K}_{\ell}\sim\frac{\beta}{\lambda_{\ell}^{1/3}}\left(1+\frac{1}{4\lambda_{\ell}}\right)+\mathcal{O}(\lambda_{\ell}^{-2})\,,
\]
with $\beta=6^{\frac{1}{3}}\Gamma\left(\frac{2}{3}\right)/\Gamma\left(\frac{1}{3}\right)\approx0.918$,
is an asymptotic approximation of $\mathcal{K}_{\ell}\equiv-\left(\ln K_{\text{i}\lambda_{\ell}}\right)'(\lambda_{\ell})$
for large $\lambda_{\ell}$; these results are derived in~\ref{subsec:Bessel--of-imaginary-1}.
Equating~\eqref{eq:LHS-deep-st} to~\eqref{eq:RHS-deep-st} and
solving in order to $\delta\epsilon_{1}$, yields 
\begin{align}
\frac{E_{p=0,\ell}}{\hbar\omega} & =\left(\ell+1\right)\left(1-\Omega_{\ell}\right)\nonumber \\
 & +\left(\frac{\gamma_{\alpha}^{2}+\ell^{2}}{2\gamma_{\alpha}}-\frac{\Delta_{\ell}-\lambda_{\ell}\mathcal{K}_{\ell}}{\gamma_{\alpha}\mathcal{K}_{\ell}^{2}}\right)\Omega_{\ell}\label{eq:spect-deep-st}
\end{align}
with 
\begin{equation}
\Omega_{\ell}=\left(1+\frac{1}{\beta^{2}}\frac{\lambda_{\ell}^{5/3}}{\lambda_{\ell}+1/2}\frac{(\ell,\gamma_{\alpha})!}{\gamma_{\alpha}^{\ell+1}e^{-\gamma_{\alpha}}}\right)^{-1}\,,\label{eq:BigOmega_ell}
\end{equation}
and 
\begin{equation}
\omega=\frac{\alpha}{2}\frac{\hbar}{m_{2}\hat{\xi}^{2}}\,,\quad\hat{\xi}^{2}=\frac{\hbar^{2}}{2n_{0}\sqrt{g_{11}m_{1}g_{12}m_{2}}}\,.\label{eq:deep-state-params}
\end{equation}
In particular, for $\ell=0$, Eq.~\eqref{eq:spect-deep-st} gives
the energy of the ground state, with~\eqref{eq:BigOmega_ell} reducing
to 
\begin{equation}
\Omega_{0}=\left(1+\frac{\gamma_{\alpha}^{2/3}}{\beta^{2}}\frac{e^{\gamma_{\alpha}}-1}{\gamma_{\alpha}+1/2}\right)^{-1}\,,
\end{equation}
for $\gamma_{\alpha}>1$, i.e. whenever it is a deep state.

We can then write down the elements of the bound-state spectrum in
the following way: 
\[
E_{p,\ell}=\begin{cases}
n_{0}g_{12}\mathcal{E}_{p,\ell}^{(\text{s})} & ,\,p-N_{\text{d}}^{(\ell)}\geq0\\
\hbar\omega\mathcal{E}_{p,\ell}^{(\text{d})} & ,\,p-N_{\text{d}}^{(\ell)}<0
\end{cases}\,,
\]
where $\mathcal{E}_{p,\ell}^{(\text{s})}$ is given by the RHS of
Eq.~\eqref{eq:bound-st-spect-ell>0} and $\mathcal{E}_{p=0,\ell}^{(\text{d})}$
by the RHS of Eq.~\eqref{eq:spect-deep-st}, and with $N_{\text{d}}^{(\ell)}$
is determined for each $\gamma^{2}$ according to~\eqref{eq:num-dpst-criterion}
and the accuracy of the deep state approximation.
\begin{figure}[t]
\includegraphics{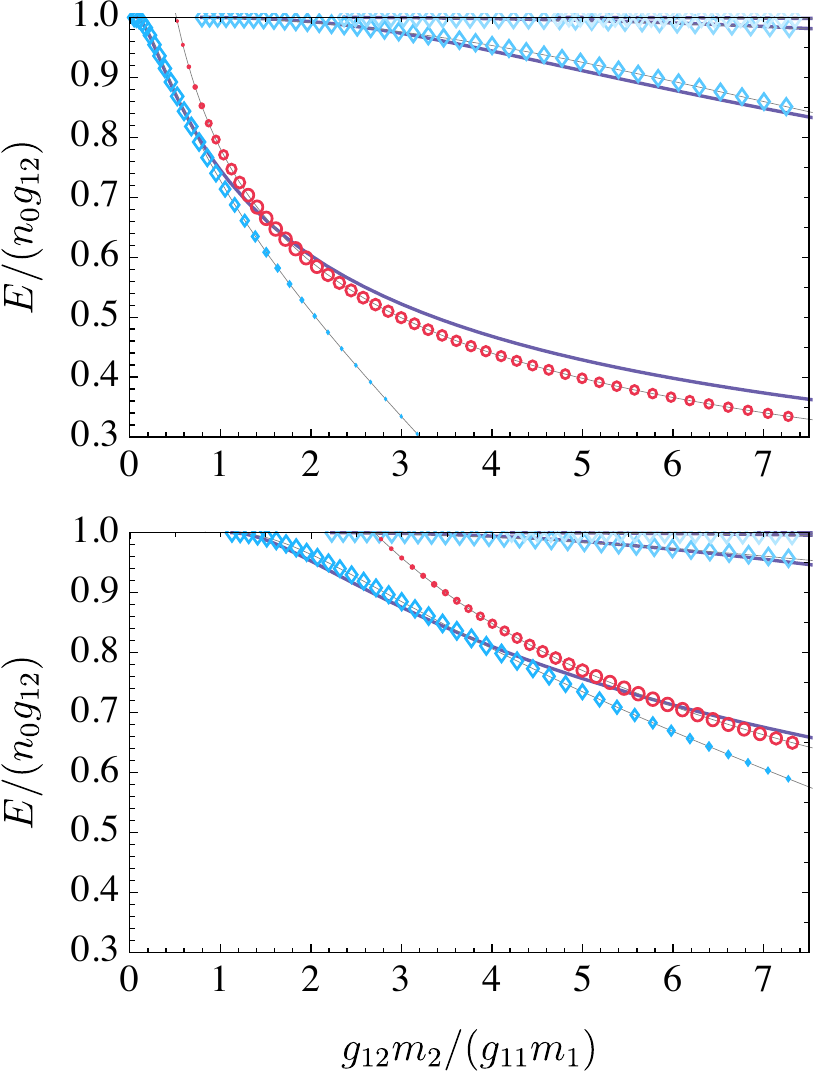}

\caption{(color online) Comparison of approximate to numerical results for
$\ell=0$ (upper panel) and $\ell=1$ (lower panel). Each group of
curves corresponds to a value of the radial quantum number $p$, starting
from $p=0$; thick solid purple lines give the numerical solution;
thin solid gray lines marked with blue diamonds (red circles) give
the solutions from the shallow-state (deep-state) approximation, with
the size of the markers being a function of the difference of approximate
and numerical solutions for equal $p$ at each $\gamma^{2}$. The
finite-size cut-off corresponds to a trap of radius $R=1000$ healing
lengths, according to Eq.~\eqref{eq:eff-bound-st-criter}, while
the horizontal axis ranges within the interval~\eqref{eq:depth-range-1}.\label{fig:Comparison-of-approximate}}
\end{figure}

\subsection{Features of the energy spectrum}

The energies of shallow and deep bound state are explicitly associated
to two different energy scales $n_{0}g_{12}$ and $\hbar\omega$;
while the former is directly sourced from the microscopic dynamics
of the system, being proportional to the interspecies coupling $g_{12}$,
the latter can be written in terms of an emergent length scale that
is \emph{not} the characteristic healing length of the condensate,
but rather a geometric mean 
\[
\hat{\xi}=\sqrt{\xi\zeta}\,,\quad\zeta^{2}=\frac{\hbar^{2}}{2n_{0}m_{2}g_{12}}\,,
\]
where $\zeta$ can be interpreted as an impurity-to-BEC penetration
length. This interpretation becomes clearer by writing the square
root of the dimensionless potential-well depth $\gamma^{2}$ as
\[
\gamma=\sqrt{\frac{g_{12}m_{2}}{g_{11}m_{1}}}=\frac{\xi}{\zeta}\,,
\]
and we see that an increase in depth of the potential can be interpreted,
instead, as increased confinement according to the two emergent length
scales. This also implies that, at least formally, $\gamma^{2}$ may
increase arbitrarily while $\hat{\xi}$, as well as $\omega$, remain
finite. As depicted in Fig.~\ref{fig:Comparison-of-approximate},
this disparity of scales indicates that, for each $\ell$, there can
be a sizeable gap between deep states and shallow states relative
to the gap between shallow states of different $p$.

Figure~\ref{fig:Comparison-of-approximate} shows excellent agreement
between the approximate and numerical solutions, though there is a
noticeable divergence with increasing $\gamma$. This is rooted in
the variational approximation~\eqref{eq:piecewise-var}: notice that
Eq.~\eqref{eq:spect-deep-st} tends to the spectrum of the harmonic
oscillator with increasing $\gamma$, accordingly with the short-range
behavior of the exact vortex profile~\eqref{eq:asymp-r-to-0}. However,
the slope at $r=0$ of the approximate solution~\eqref{eq:piecewise-var}
is smaller than that of the exact one, as fixed by the variational
procedure. This discrepancy thus becomes starker with increased depth,
with the spectrum of deep states being directly proportional to this
slope as shown in Eq.~\eqref{eq:deep-state-params}.

\begin{figure}[t]
\includegraphics{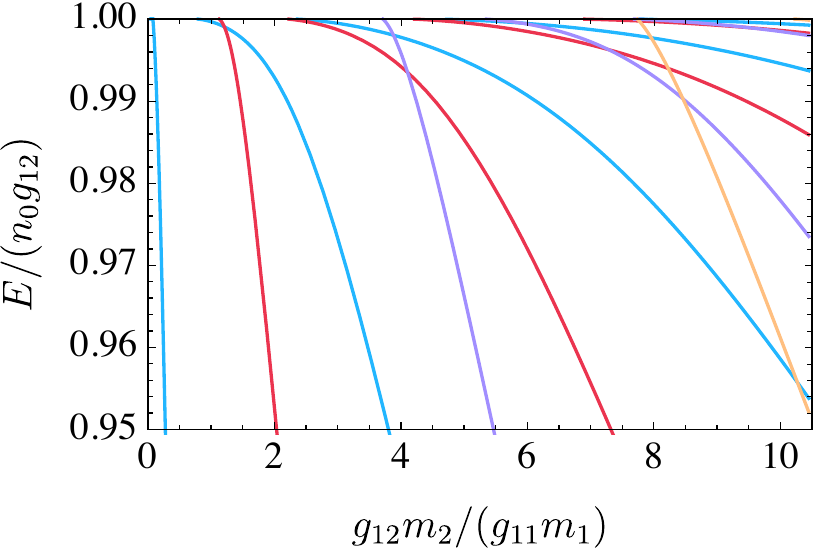}\caption{(color online) Numerical values of the spectrum as a function of $\gamma^{2}$
fpr $\ell=0$ (blue), $\ell=1$ (red), $\ell=2$ (purple) and $\ell=3$
(orange). Energy levels have a finite-size cut-off corresponding
to a trap of radius $R=1000$ healing lengths.\label{fig:Spectrum-as-a}}
\end{figure}

An interesting feature of the impurity spectrum has to do with the
energy levels near the edge of the potential. In this region, the
spectrum is rich in crossovers and accidental degeneracies, resulting
from the visibly faster decrease in energy of states of higher angular
momenta. This can be understood in light of the competition between
the centrifugal barrier and the vortex potential: at the onset of
the first bound states, the long-range profile of the potential is
canceled by the centrifugal term, rendering the effective potential
more confining than that of states of lower angular momenta. This behavior is displayed in Fig.~\eqref{fig:Spectrum-as-a}.

\section{Finite size and physical bound states\label{sec:Finite-size-effects-and}}

The spectrum obtained in Eq. \eqref{eq:bound-st-spect-ell>0} for
the shallow states indicates that there are, in fact, infinitely many
bound-state solutions for each $\ell<\gamma_{\alpha}$, which can
be attributed to the slow algebraic growth of the vortex profile.
Most of these are not physical solutions, however. To see this, we must consider that there is a natural cut-off imposed
by the finite size of the trap, which is not merely a practical limitation:
this is required by the stability of the vortex configuration, whose
energy diverges logarithmically with the system size~\cite{Lundh:1998aa},
and also due to the impossibility of long-range order in low dimensions,
as imposed by the Mermin-Wagner theorem. We can account for this simply
by requiring that the classical turning point (in the radial coordinate)
of the putative bound state is lying within the domain of the BEC.
Hence, assuming a disk-like trap of radius $R\gg\xi$ , we have the
quantitative condition 
\begin{equation}
1-\frac{E_{p,\ell}}{n_{0}g_{12}}>\frac{\xi^{2}}{\alpha^{2}R^{2}}\,,\label{eq:eff-bound-st-criter}
\end{equation}
which, together with Eq. \eqref{eq:bound-st-spect-ell>0}, imposes
a condition in $\gamma^{2}$ for a bound state $(p,\ell)$ to be contained
in the BEC. Fig.~\eqref{fig:Comparison-of-approximate} shows comparisons
between numerical and approximate solutions, where the cut-off \eqref{eq:eff-bound-st-criter}
for a trap of size $R=1000\xi$ was included. It is clear that for
each value of $\gamma$ there are only a finite number of bound states
of the vortex, while those violating the condition Eq. \eqref{eq:eff-bound-st-criter}
can be seen as bound states of the trapping potential.

\begin{figure}
\flushleft
\includegraphics{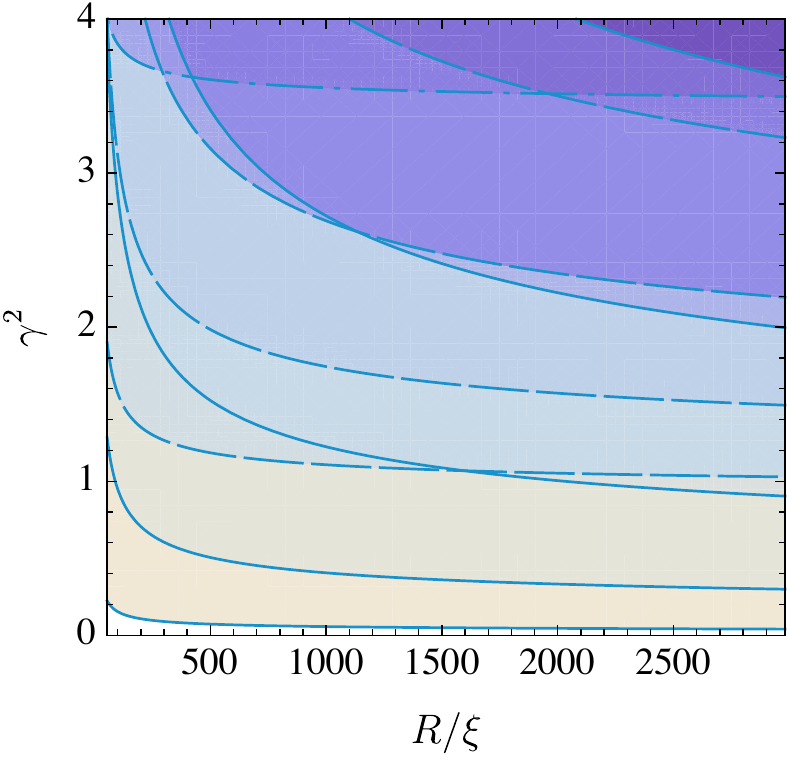}
\caption{(color online) Diagram of regimes of the vortex potential well, with the critical lines of~\eqref{eq:p0-condition-boundst} for
$\ell=0$ (solid), $\ell=1$ (dashed) and $\ell=2$ (dot-dashed) bound states.
For each point $(R/\xi,\gamma^{2})$ on the plane, the number of bound
states is given by the number of curves below it, with a multiplicity of 2 for $\ell\geq1$ (e.g. the point $(1000,2.4)$
has $3+2\times2=7$ bound states). Darker shades correspond to greater number of bound states.}
\label{fig:Diagram-regimes} 
\end{figure}

Solutions to Eq.~\eqref{eq:eff-bound-st-criter} can be found semi-analytically,
as discussed in~\ref{subsec:Onset-of-bound}, yielding the critical
condition
\begin{equation}
\left.\frac{g_{12}m_{2}}{g_{11}m_{1}}\right|_{p,\ell}>\alpha^{2}\ell^{2}+\frac{c_{p,\ell}^{-2}}{\ln\left(\frac{R}{r_{p,\ell}\xi}\right)^{2}}
\label{eq:p0-condition-boundst}
\end{equation}
for the onset of the bound state $(p,\ell)$. The coefficients $c_{p,\ell}$
and $r_{p,\ell}$ for the first few bound states are given in Table~\eqref{tab:Results-of-the}.
Figure~\ref{fig:Diagram-regimes} gives a diagram of regimes of the vortex potential
well, providing the number of bound states for each point in the $(R/\xi,\gamma^{2})$
parameter space.

Moreover, we can expect that imposing this cut-off \emph{a posteriori}
has negligible effects on the physical solutions of Eq.~\eqref{eq:Schr-prob}
since, in practice, the condition $R\gg\xi$ is implied by the thermodynamic
limit. The caveat is that the possible highly-excited, highly-delocalized
states may become affected by the trapping potential, but this correction
can, in principle, be accounted for using perturbation theory. For
this reason, the RHS of Eq.~\eqref{eq:p0-condition-boundst} is a
tight lower bound of the critical value for the onset of the bound
state $(p,\ell)$ .

\section{Experimetnal considerations\label{sec:Experimetnal-considerations}}

Recent experimental work with Yb-$^{7}\text{Li}$ mixtures~\cite{Schafer:2018aa}
provides a realistic platform for the realization of this system:
by virtue of the substantial mass imbalance of $m_{\text{Yb}}/m_{\text{Li}}\approx25$,
the impurity-to-vortex decoupling condition in Eq. \eqref{eq:decoup-condition}
is within reach by immersion of Yb impurities (either fermionic $^{173}\text{Yb}$,
or bosonic $^{174}\text{Yb}$~\cite{Schafer:2018aa}) in a $^{7}\text{Li}$
condensate, while also increasing $\gamma^{2}$.

Nevertheless, the tunability of the latter is still necessary, since
a vortex supporting only a few bound states requires $\gamma^{2}$
one order of magnitude below that; thus, Feshbach resonances would
be a crucial element in this realization. Coincidentally, the realizations
of Ref.~\cite{Schafer:2018aa} exploits a fairly accessible Feshbach
resonance of $^{7}\text{Li}$ in order to produce a stable and sizeable
BEC of this species~\cite{Gross:2008aa}. The modest interspecies
scattering found for Yb-$^{7}\text{Li}$ ($\sim1\,\text{nm}$)~\cite{Schafer:2018aa}
then means that the parameter range in Eq. \eqref{eq:depth-range-1}
is well within reach by tuning the $^{7}\text{Li}$ intra-species
scattering to tens of nanometers.

In a quasi-2D BEC, such as considered presently, couplings become
renormalized by the transversal length scale $a_{\perp}$ of the harmonic
trap~\cite{salasnich_2002,krueger_2007,young_2010}, as $g_{ij}\propto a_{ij}/a_{\perp}$
if $a_{ij}$ is the scattering length between species $i$ and $j$.
While this multiplicative factor has no effect on $\gamma^{2}\propto g_{12}/g_{11}$,
the decoupling parameter $\kappa^{2}/\gamma^{2}$ (see Eq.~\eqref{eq:decoup-condition})
is proportional to $g_{11}$ alone. Typical trapping frequencies of
tens of kHz yield a transversal length of hundreds of nanometers for
$^{7}$Li, which further reduces the decoupling parameter \cite{zhang_2008}.
Additional flexibility comes from the usage of box potentials \cite{gaunt_2013,gotlibovych_2014},
which are able to produce homogeneous condensates in a more controllable
way \cite{Chomaz_2015,desbuquois_2014}.

Another important aspect pertains to the relation of the energy scale
$n_{0}g_{12}$ of the spectrum to the temperature. We find that for
a peak density (in volume) of $\sim10^{4}\,\text{cm}^{-3}$, the Yb-$^{7}\text{Li}$
mixture yields $n_{0}g_{12}/k_{\text{B}}\propto10^{2}\,\text{nK}$,
meaning that the typical energy gap is much larger than the temperature
of the system. This suggests that we might be in good position to
further exploit these impurity bound states as physical qubits \cite{Shaukat:2017aa},
thus paving the stage for a possible quantum information platform
operating in the kHz range.

\section{Conclusion\label{sec:Conclusion}}

Starting from a variational ansatz for the vortex profile in a quasi
two-dimensional Bose-Einstein condensate, which has been shown to
be in excellent agreement with the numerical calculations, we have
obtained the eigenstates and the eigenvalues of the vortex-impurity
problem. Our method consists in imposing regularity conditions (continuity
and finite derivative) at the crossing of the vortex piecewise solution,
the point marking the transition between the core ($\sim r$, as $r\rightarrow0$)
and the edge ($\sim1/r^{2}$, as $r\rightarrow\infty$) of the vortex.
As a result, we are able to obtain analytic expression for both shallow
and deep bound states, respectively lying at the edge and at the bottom
of the effective potential experienced by the impurity when interacting
with the vortex. A comparison with numerical results reveals that,
for the states of angular momentum $\ell=0,1$, our analytical results
appear to be accurate in their respective range of validity. For the
general case, for states lying at the vortex profile transition, analytical
expressions are, in general, not available explicitly. We point out
that similar results have been obtained in the study of the one-dimensional
$1/x^{2}$ potential~\cite{Essin:2006aa}, and that a similar approach
was used in the study of core-to-coreless vortex transition in multicomponent
superfluids \cite{Catelani:2010aa}. 

The consideration of heavy impurities immersed in a gas of light particles,
as made possible by the recent experiments allowing for the controllable
mixture of Li and Yb \cite{hansen_2013,khramov_2014,Schafer:2018aa},
allows the investigation of the polaron physics in a fashion opposed
to the usual solid-state scenarios, where the impurity (electron)
is much lighter that the host particles (ions) \cite{frohlich_1950}.
Moreover, our calculations may also contribute for future studies
of the so-called ``Tkachenko polaron\char`\"{}, a quasi-particle
resulting from the coupling between an impurity and a vortex lattice
vibration in rotating Bose-Einstein condensates \cite{caracanhas_2013}.
Here, the vortex-induced trapping may significantly change the features
of the polaron.

More crucially, our findings show that it is possible to tune the
value of the impurity-BEC interaction to isolate two deep (localized)
bound states deep in the core of the vortex, making it possible to
promote the impurity into a qubit, an essential element for applications
in quantum information theory. In the future, it will be crucial to
investigate the relevant beyond mean-field effects, namely the coupling
with the quantum excitations of the BEC (phonons), a task that we
believe essential for the complete characterization of the qubit performance.
Due to the scalability of the number of vortices in a two-dimensional
BEC, we expect this to become an interesting alternative to the one-dimensional
qubits made of dark-solitons \cite{Shaukat:2019aa}, therefore offering
a possible alternative for a quantum information platform operating
with acoustic degrees of freedom in a near future. \\
 
\begin{acknowledgments}
The authors acknowledge the financial support of FCT-Portugal through
grant No. PD/BD/128625/2017 and through the contract number IF/00433/2015.
JEHB would also like to thank the Joint Quantum Institute and the
University of Maryland at College Park, for their hospitality during
the writing of this manuscript, as well as the Fulbright Commission
in Portugal. HT further acknowledges financial support from the Quantum
Flagship Grant PhoQuS (820392) of the European Union. 
\end{acknowledgments}

\appendix

\section{Properties of the confluent hypergeometric function \label{sec:Leading-order-of}}

\subsection{Zeros and monotonicity\label{subsec:Zeros-and-monotonicity}}

The confluent hypergeometric function (CHF) can be expressed as a
generalized hypergeometric series, 
\begin{equation}
M(a,b;x)=\sum_{m=0}^{\infty}\frac{(a)_{m}}{(b)_{m}}\frac{x^{m}}{m!}\,,\label{eq:chf-series}
\end{equation}
where 
\begin{equation}
(a)_{m}=\frac{\Gamma(a+m)}{\Gamma(a)}=a(a+1)\text{\ensuremath{\dots}}(a+m-1)\,,\label{eq:pochhammer}
\end{equation}
for any $a\in\mathbb{C}$, is the Pochhammer symbol, also known as
the rising factorial; the CHF is an entire function of its argument,
implying that the series~\eqref{eq:chf-series} has an infinite radius
of convergence.~\cite{Olver:2010:NHM:1830479}

From this definition, it is clear that for $x\in[0,+\infty[$ the
CHF has no zeros for any real $a,b\geq0$ and that it is positive
and increasing.~\cite{Olver:2010:NHM:1830479} Moreover, when $a<0$
it has $\lceil-a\rceil$ positive zeros, and when $a=-n$, with $n=0,1,2,\text{\ensuremath{\dots}}$,
and $b=\beta+1$, with $\beta$ real and non-negative, the CHF reduces
to the $n$th generalized Laguerre polynomial, 
\begin{equation}
M(-n,\beta+1;x)=\frac{n!}{(\beta+1)_{n}}L_{n}^{(\beta)}(x)\,;\label{eq:chf-to-Lag}
\end{equation}
the series in~\eqref{eq:chf-series} reduces to an $n$th-degree
polynomial since $(-n)_{m}$ vanishes for $m\geq n-1$, as per~\eqref{eq:pochhammer}.~\cite{Olver:2010:NHM:1830479}
In particular, the generalized Laguerre polynomials comprise the solutions
of the isotropic quantum harmonic oscillator.~\cite{Olver:2010:NHM:1830479}

\subsection{Derivative with respect to the first parameter\label{subsec:Derivative-with-respect}}

Suppose that in Eq.~\eqref{eq:chf-series} we have $a=-n+\delta a$
with $n=0,1,2,\text{\ensuremath{\dots}}$, with $\delta a$ real and
of small absolute value, and $b=\beta+1$, with $\beta$ real and
non-negative. Taylor-expanding with regard to the first parameter
we have 
\begin{align}
 & M(-n+\delta a,\beta+1;x)=\nonumber \\
 & =\frac{n!}{(\beta+1)_{n}}L_{n}^{(\beta)}(x)+\delta a\,\tilde{M}_{n}^{(\beta)}(x)+O(\delta a^{2})\,,\label{eq:a-expand-chf}
\end{align}
where we use~\eqref{eq:chf-to-Lag} for the zeroth-order term, and
define 
\begin{equation}
\tilde{M}_{n}^{(\beta)}(x)=\left.\frac{\partial}{\partial a}M\left(a,\beta+1;x\right)\right|_{a=-n},\label{eq:a-1deriv-chf}
\end{equation}
for $n=0,1,2,\text{\ensuremath{\dots}}$. For $n=0$, we can use Eq.~(38a)
of Ref.~\cite{Ancarani:2008aa} and rewrite it as 
\begin{equation}
\tilde{M}_{0}^{(\beta)}(x)=\frac{1}{\beta+1}\sum_{m=0}^{\infty}\frac{x^{m+1}}{(m+1)(\beta+2)_{m}}\,;\label{eq:a0-deriv1-chf-series}
\end{equation}
notice that 
\[
\frac{\text{d}}{\text{d}x}\tilde{M}_{0}^{(\beta)}(x)=\frac{1}{\beta+1}M(1,\beta+2;x)\,,
\]
since $(1)_{m}=m!$ according to~\eqref{eq:pochhammer}; this CHF
can be written as~\cite{Olver:2010:NHM:1830479} (p. 328, Eq.~(13.6.5))
\begin{equation}
M(1,\beta+2;x)=(\beta+1)x^{-\beta-1}e^{x}(\beta,x)!\,,\label{eq:chf-to-incgamma}
\end{equation}
where $(\beta,x)!$ is a rewriting of the incomplete gamma function~\cite{footnote_1},
\begin{equation}
(\beta,x)!=\gamma(\beta+1,x)=\int_{0}^{x}\text{d}t\,t^{\beta}e^{-t}\,,\label{eq:incgamma-notations}
\end{equation}
so that we can write~\eqref{eq:a0-deriv1-chf-series} as an integral
of~\eqref{eq:chf-to-incgamma}: 
\begin{equation}
\tilde{M}_{0}^{(\beta)}(x)=\int_{0}^{x}\frac{\text{d}y}{y}\frac{(\beta,y)!}{y^{\beta}e^{-y}}\,.\label{eq:a0-1deriv-chf-int}
\end{equation}
Further, for $\beta=\ell$ a non-negative integer, it follows from~\eqref{eq:incgamma-notations}
that~\cite{Olver:2010:NHM:1830479} (p. 177, Eq.~(8.4.7))
\begin{equation}
\frac{(\ell,x)!}{\ell!}=1-e^{-x}\sum_{k=0}^{\ell}\frac{x^{k}}{k!}\,.\label{eq:inc-gamma-integer}
\end{equation}
In turn, for $n\geq1$ we have \cite{Olver:2010:NHM:1830479} (p. 325, Eq.~(13.3.1)) 
\begin{equation}
 M(a,b;x)=\frac{(a-b)M(a-1,b;x)}{2a-b+x}+\frac{aM(a+1,b;x)}{2a-b+x}\,;
\end{equation}
identifying parameters, expanding both sides to first order in $\delta a$
and matching powers, we find 
\begin{widetext}
\begin{align}
n=1:\quad & \left(\beta+1\right)\tilde{M}_{1}^{(\beta)}(x)=-\frac{x}{\beta+1}+\frac{\left(\beta,x\right)!}{x^{\beta}e^{-x}}+\left(1+\beta-x\right)\int_{0}^{x}\frac{\text{d}y}{y}\frac{(\beta,y)!}{y^{\beta}e^{-y}}\,,\\
n\geq2:\quad & \left(\beta+n\right)\tilde{M}_{n}^{(\beta)}(x)-(2n+\beta-x-1)\tilde{M}_{n-1}^{(\beta)}(x)+(n-1)\tilde{M}_{n-2}^{(\beta)}(x)=\sum_{k=0}^{2}\binom{2}{k}\frac{(-1)^{k}(n-k)!}{(\beta+1)_{n-k}}L_{n-k}^{(\beta)}(x)\,;
\end{align}
\end{widetext}
for $n\geq2$ we use~\eqref{eq:chf-to-Lag} to write the RHS in terms
of generalized Laguerre polynomials, while for $n=1$ we use~\eqref{eq:chf-to-Lag},
and~\eqref{eq:chf-to-incgamma} along with the recurrence relation
of the incomplete gamma function~\cite{Olver:2010:NHM:1830479} (p.
178, Eq.~(8.8.1)) 
\begin{equation}
\beta\left(\beta-1,x\right)!=x^{\beta}e^{-x}+\left(\beta,x\right)!\,.
\end{equation}

\section{Asymptotics of the Bessel-$K$ of imaginary order and its derivatives}

\subsection{Bessel-$K$ of imaginary order for small argument\label{subsec:Bessel--of-imaginary}}

For $x\in\mathbb{R}^{+}$ small and $\lambda\in\mathbb{R}^{+}$, we
have the limiting behavior~\cite{Olver:2010:NHM:1830479,Dunster:1990aa}
\begin{equation}
K_{\text{i}\lambda}(x)=-\sqrt{\frac{\pi}{\lambda\sinh(\pi\lambda)}}\sin\left(\lambda\ln\frac{x}{2}-\varphi(\lambda)\right)+\mathcal{O}(x^{2})\,,\label{eq:small-arg-K}
\end{equation}
where the function $\varphi(\lambda)$ is given by 
\begin{equation}
\varphi(\lambda)=\arg\Gamma(1+\text{i}\lambda)\,,\label{eq:phi_arg_Gamma}
\end{equation}
with the branch defined so that $\varphi$ is continuous for $0<\lambda<\infty$
and $\lim_{\lambda\rightarrow0}\varphi(\lambda)=0$~\cite{Dunster:1990aa}.
For large $\lambda$, this can be approximated using Stirling's formula~\cite{Olver:2010:NHM:1830479},
yielding 
\begin{equation}
\varphi(\lambda)\sim\frac{\pi}{4}+\lambda\left(\ln\lambda-1\right)+\mathcal{O}(\lambda^{-2})\,,\label{eq:stirling-arg-gamma}
\end{equation}
while for small $\lambda$, the gamma function can be approximated
by the series~\cite{Olver:2010:NHM:1830479} (p. 139, Eq.~(5.7.3))
\begin{align}
\log\Gamma(1+z) & =-\log(1+z)+(1-\gamma_{\text{E}})z\nonumber \\
 & +\sum_{k=2}^{\infty}(-1)^{k}\left(\zeta(k)-1\right)\frac{z^{k}}{k}\,,
\end{align}
for $\left|z\right|<2$, where $\gamma_{\text{E}}\approx0.5572$ is
the Euler-Mascheroni constant and $\zeta$ the Riemann-zeta function,
giving 
\begin{align}
\varphi(\lambda) & =-\gamma_{\text{E}}\lambda-\sum_{k=1}^{\infty}(-1)^{k}\frac{\zeta(2k+1)}{2k+1}\lambda^{2k+1}\,.\label{eq:varphi_leading}
\end{align}

\subsection{Bessel-$K$ of imaginary order at the transition point\label{subsec:Bessel--of-imaginary-1}}

In deriving Eq.~\eqref{eq:RHS-deep-st}, we perform a Taylor expansion
\begin{align}
 & \left.\frac{\text{d}\ln K_{\text{i}\lambda_{\ell}}(qr)}{\text{d}r}\right|_{r=r_{\alpha},q=\frac{\lambda_{\ell}}{r_{\alpha}}+\delta\epsilon_{2}}=\nonumber \\
 & =-\frac{\lambda_{\ell}}{r_{\alpha}}\mathcal{K}_{1}(\lambda_{\ell})-\delta\epsilon_{2}\left(\mathcal{K}_{1}(\lambda_{\ell})+\lambda_{\ell}\mathcal{K}_{2}(\lambda_{\ell})\right)+\mathcal{O}(\delta\epsilon_{2}^{2})\label{eq:pre-RHS-deep-st}
\end{align}
where we define a family of functions specified as 
\begin{equation}
\mathcal{K}_{n}(\lambda_{\ell})=-\left(\ln K_{\text{i}\lambda_{\ell}}\right)^{(n)}(\lambda_{\ell})\,,
\end{equation}
that is, the $n$th derivative of the function $-\ln K_{\text{i}\lambda_{\ell}}(x)$
evaluated at the so-called transition point $x=\lambda_{\ell}$. For
$n=1,2$, we have explicitly 
\begin{equation}
\mathcal{K}_{1}(\lambda_{\ell})=-\frac{K_{\text{i}\lambda_{\ell}}'\left(\lambda_{\ell}\right)}{K_{\text{i}\lambda_{\ell}}\left(\lambda_{\ell}\right)}\,,\label{eq:special-val}
\end{equation}
\begin{equation}
\mathcal{K}_{2}(\lambda_{\ell})=-\frac{K_{\text{i}\lambda_{\ell}}''\left(\lambda_{\ell}\right)}{K_{\text{i}\lambda_{\ell}}\left(\lambda_{\ell}\right)}+\left(\frac{K_{\text{i}\lambda_{\ell}}'\left(\lambda_{\ell}\right)}{K_{\text{i}\lambda_{\ell}}\left(\lambda_{\ell}\right)}\right)^{2}\,,\label{eq:special-val-deriv}
\end{equation}
where $\lambda_{\ell}>\sqrt{2\ell+1}$, $\ell\in\mathbb{N}_{0}$.
Note that $\mathcal{K}_{\ell}\equiv\mathcal{K}_{1}(\lambda_{\ell})$,
as defined in the main text.

We have, for any $z,\nu\in\mathbb{C}$~\cite{Olver:2010:NHM:1830479}
(p. 252, Eqs.~(10.29.2)), 
\begin{equation}
K_{\nu}'(z)=\frac{\nu}{z}K_{\nu}(z)-K_{\nu+1}(z)\,,
\end{equation}
\begin{equation}
K_{\nu}''(z)=\left(1+\frac{\nu(\nu-1)}{z^{2}}\right)K_{\nu}(z)+\frac{1}{z}K_{\nu+1}(z)\,;
\end{equation}
in particular, for $z=\lambda$ and $\nu=\text{i}\lambda$, $\lambda\in\mathbb{R}$,
we have 
\begin{align}
K_{\text{i}\lambda}'(\lambda) & =\text{i}K_{\text{i}\lambda}(\lambda)-K_{1+\text{i}\lambda}(\lambda)\,,
\end{align}
\begin{equation}
K_{\text{i}\lambda}''(\lambda)=-\frac{\text{i}}{\lambda}K_{\text{i}\lambda}(\lambda)+\frac{1}{\lambda}K_{1+\text{i}\lambda}(\lambda)=-\frac{1}{\lambda}K_{\text{i}\lambda}'(\lambda)\,,
\end{equation}
and then Eq.~\eqref{eq:special-val-deriv} can then be written in
terms of $\mathcal{K}_{\ell}$ only as 
\begin{equation}
\mathcal{K}_{2}(\lambda_{\ell})=-\frac{\mathcal{K}_{\ell}}{\lambda_{\ell}}+\mathcal{K}_{\ell}^{2}\,.\label{eq:second-deriv-lnK}
\end{equation}
Substituting~\eqref{eq:second-deriv-lnK} into~\eqref{eq:pre-RHS-deep-st},
we obtain Eq.~\eqref{eq:RHS-deep-st}.

We have reduced the problem to computing asymptotic expressions of
functions $K_{\text{i}\lambda}(\lambda)$ and $K_{\text{i}\lambda}'(\lambda)$
of $\lambda$, for $\lambda>1$. We make use of the integral representation
\cite{Olver:2010:NHM:1830479}\textcolor{black}{{} (p. 252}, Eq.~(10.32.9))
\begin{equation}
K_{\nu}(z)=\int_{0}^{\infty}\text{d}t\,e^{-z\cosh t}\cosh(\nu t)\,,\label{eq:integ-rep-BesselK}
\end{equation}
or $\left|\text{ph}z\right|<\frac{\pi}{2}$ and any $\nu\in\mathbb{C}$.
Substituting $\nu=\text{i}\lambda$ and $z=\lambda$, we have 
\begin{align}
K_{\text{i}\lambda}(\lambda) & =\int_{0}^{\infty}\text{d}t\,e^{-\lambda\cosh t}\cos(\lambda t)\nonumber \\
 & =\frac{1}{2}\int_{-\infty}^{+\infty}\text{d}t\,e^{-\lambda\cosh t+\text{i}\lambda t}\,,
\end{align}
and we find 
\begin{equation}
K_{\text{i}\lambda}'(\lambda)=\frac{\text{d}K_{\text{i}\lambda}(\lambda)}{\text{d}\lambda}-\frac{1}{2}\int_{-\infty}^{+\infty}\text{d}t\,\text{i}te^{-\lambda\cosh t+\text{i}\lambda t}\,.
\end{equation}
We can thus obtain asymptotic forms of $K_{\text{i}\lambda}(\lambda)$
and $K_{\text{i}\lambda}'(\lambda)$ from two $\lambda$-dependent
integrals 
\begin{equation}
\mathcal{I}_{k}=\frac{1}{2}\int_{-\infty}^{+\infty}\text{d}t\,\left(\text{i}t\right)^{k}e^{-\lambda\cosh t+\text{i}\lambda t}\,,\quad k=0,1\,.
\end{equation}

\begin{figure}[t]
\includegraphics{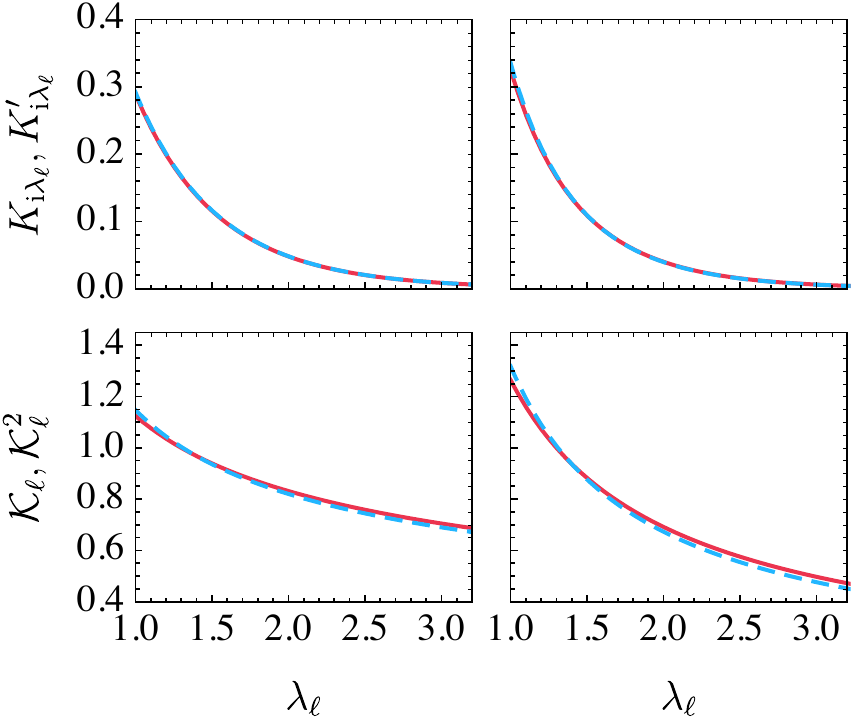}\caption{Comparison of asymptotic approximations (dashed blue) and exact functions
(solid red); top-right $K_{\text{i}\lambda_{\ell}}$, top-left $K_{\text{i}\lambda_{\ell}}'$,
bottom-right $\mathcal{K}_{\ell}$ and bottom-left $\mathcal{K}_{\ell}^{2}$.
The horizontal axis ranges within the interval~\eqref{eq:depth-range-1}
for $\lambda_{\ell}>1$, for each $\ell$.\label{fig:Comparison-of-asymptotic}}
\end{figure}

We note that $\cosh t=2\sinh^{2}\frac{t}{2}+1$ and transform the
integration coordinate as 
\[
\sinh\frac{t}{2}=\frac{x}{2}\Rightarrow\text{dt}=\frac{\text{d}x}{\sqrt{1+\left(\frac{x}{2}\right)^{2}}}\,,
\]
yielding 
\begin{equation}
\mathcal{I}_{k}=\frac{e^{-\lambda}}{2}\int_{-\infty}^{+\infty}\text{d}x\,\frac{e^{-\lambda g(x)}}{\sqrt{1+\left(\frac{x}{2}\right)^{2}}}\left(\text{i}t(x)\right)^{k}\,,\quad k=0,\,1\,,\label{eq:stat-phase_form-1}
\end{equation}
with $t(x)=2\text{i}\sinh^{-1}\frac{x}{2}$ and $g(x)=\frac{x^{2}}{2}-2\text{i}\sinh^{-1}\frac{x}{2}$.
We now perform a stationary phase approximation~\cite{Olver_Asymp_1997}:
the function $g$ is stationary for $x=x_{*}=\text{i}\sqrt{2}$ and,
expanding to subleading order in $\xi=x-x_{*}$, we have 
\begin{equation}
g(x_{*}+\xi)=\left(\frac{\pi}{2}-1\right)+\frac{\text{i}\sqrt{2}}{3}\xi^{3}+\frac{\xi^{4}}{2}+\mathcal{O}(\xi^{5})\,.\label{eq:stationary-g}
\end{equation}
Here, we include the subleading term to make a point that the integral
is convergent; in what follows, however, we include only the leading
order in $\xi$ in the exponent, meaning we assume that contributions
to the integral~\eqref{eq:stat-phase_form-1} result dominantly from
the oscillatory factor rather than from the decaying one, the latter
granting only (implicitly) the convergence of the integral. We then
have 
\begin{equation}
\mathcal{I}_{k}\sim\frac{e^{-\frac{\pi\lambda}{2}}}{2}\int_{-\infty}^{+\infty}\text{d}\xi\,\frac{e^{-\frac{\text{i}\sqrt{2}\lambda}{3}\xi^{3}}}{\sqrt{1+\left(\frac{x_{*}+\xi}{2}\right)^{2}}}p_{k}(\xi)\,,
\end{equation}
where we have made $p_{k}(\xi)=\left(\text{i}t(x_{*}+\xi)\right)^{k}$
, so that $p_{0}(\xi)=1$ and 
\begin{equation}
p_{1}(\xi)=-\frac{\pi}{2}+\sqrt{2}\text{i}\xi-\frac{1}{2}\left(\text{i}\xi\right)^{2}+\frac{\sqrt{2}}{3}\left(\text{i}\xi\right)^{3}-\frac{1}{2}\left(\text{i}\xi\right)^{4}+\mathcal{O}(\xi^{5})\,.
\end{equation}
Moreover, we consider that the variation of the denominator of the
integrand is negligible in the presence of the exponential factor,
so that we have 
\begin{equation}
\int_{-\infty}^{+\infty}\frac{\text{d}\xi\,e^{-\frac{\text{i}\sqrt{2}\lambda}{3}\xi^{3}}}{\sqrt{1+\left(\frac{x_{*}+\xi}{2}\right)^{2}}}p_{k}(\xi)\approx\sqrt{2}\int_{-\infty}^{+\infty}\text{d}\xi\,e^{-\frac{\text{i}\sqrt{2}\lambda}{3}\xi^{3}}p_{k}(\xi)\,,
\end{equation}
and, then, 
\begin{align}
\frac{1}{2}\int_{-\infty}^{+\infty}\text{d}x\,e^{-\frac{\text{i}\sqrt{2}\lambda}{3}\xi^{3}}p_{k}(\xi) & =\text{Re}\left(\int_{0}^{\infty}\text{d}\xi\,e^{-\frac{\text{i}\sqrt{2}\lambda}{3}\xi^{3}}p_{k}(\xi)\right)\,,\label{eq:third-order-pseudo-gaussian-1-1}
\end{align}
since $p_{k}(-\xi)$ is the complex conjugate of $p_{k}(\xi)$. We
have the integral identity in Ref.~\cite{Jeffrey:2007aa}
(p. 337, Eq.~(3.326.2.)) 
\begin{equation}
\int_{0}^{\infty}\text{d}x\,e^{-\beta x^{n}}x^{m}=\frac{\Gamma(\mu)}{n\beta^{\mu}}\,,\quad\mu=\frac{m+1}{n}\,,\label{eq:integration-saddle-pt-1}
\end{equation}
for $\text{Re}\beta,\,\text{Re}m,\,\text{Re}n>0$ \cite{footnote_2}.
Thus, for the integral of~\eqref{eq:third-order-pseudo-gaussian-1-1},
we will have 
\begin{align}
 & \text{Re}\left(\sqrt{2}\int_{0}^{\infty}\text{d}\xi\,e^{-\frac{\text{i}\sqrt{2}\lambda}{3}\xi^{3}}\left(\text{i}\xi\right)^{m}\right)\nonumber \\
 & =\frac{1}{3}\cos\left(\frac{\pi}{2}\frac{2m-1}{3}\right)\frac{6^{\frac{m+1}{3}}\Gamma\left(\frac{m+1}{3}\right)}{2^{\frac{m}{2}}\lambda^{\frac{m+1}{3}}}\,,
\end{align}
and we arrive at

\begin{equation}
K_{\text{i}\lambda}(\lambda)=\mathcal{I}_{0}\sim\frac{e^{-\frac{\pi\lambda}{2}}}{2\sqrt{3}}\frac{6^{\frac{1}{3}}\Gamma\left(\frac{1}{3}\right)}{\lambda^{\frac{1}{3}}}\,,\label{eq:bessK-asym}
\end{equation}

\begin{equation}
K_{\text{i}\lambda}'(\lambda)=\frac{\text{d}\mathcal{I}_{0}}{\text{d}\lambda}-\mathcal{I}_{1}\sim-\frac{e^{-\frac{\pi\lambda}{2}}}{2\sqrt{3}}\frac{6^{\frac{2}{3}}\Gamma\left(\frac{2}{3}\right)}{\lambda^{\frac{2}{3}}}\left(1+\frac{1}{4\lambda}\right)+\mathcal{O}(\lambda^{-2})\,,\label{eq:deriv-bessK-asymp}
\end{equation}
and, moreover, 
\begin{equation}
\mathcal{K}_{\ell}=-\frac{K_{\text{i}\lambda_{\ell}}'\left(\lambda_{\ell}\right)}{K_{\text{i}\lambda_{\ell}}\left(\lambda_{\ell}\right)}\sim\frac{\kappa}{\lambda_{\ell}^{1/3}}\left(1+\frac{1}{4\lambda_{\ell}}\right)+\mathcal{O}(\lambda_{\ell}^{-2})\,,
\end{equation}
with $\kappa=6^{\frac{1}{3}}\Gamma\left(\frac{2}{3}\right)/\Gamma\left(\frac{1}{3}\right)\approx0.918$.
Fig.~\ref{fig:Comparison-of-asymptotic} shows a comparison of the
exact functions and respective asymptotic approximations. Moreover,
these appear to be in agreement with the formulas provided in Ref.~\cite{Magnus:2013aa}
(p. 142, 2nd eq.).

\section{Onset of bound states}

\subsection{Nonexistence of bound states of angular momenta $\ell\geq\gamma_{\alpha}$\label{sec:Inexistence-of-bound}}

We show that whenever $\Delta_{\ell}=\gamma_{\alpha}-\ell\leq0$,
Eq.~\eqref{eq:log-deriv} has no solutions for $q$ real and positive.

We note that $K_{\nu}(x)$ of real order $\nu$ is a positive and
decreasing function of $x\in]0,+\infty[$, hence we have 
\begin{equation}
\left(\ln R_{2}\right)'(r_{\alpha})=q\frac{K_{\nu}'(qr_{\alpha})}{K_{\nu}(qr_{\alpha})}<0\label{eq:bound-BesselK}
\end{equation}
for all $q>0$, where $\nu=\sqrt{\ell^{2}-\gamma_{\alpha}^{2}}$.
It follows that solutions of~\eqref{eq:log-deriv} would require
\begin{equation}
\left(\ln R_{1}\right)'(r_{\alpha})<0\,.\label{eq:cond-reg1}
\end{equation}
In turn, according to the properties of the CHF for positive parameters
presented in Appendix~\ref{subsec:Zeros-and-monotonicity}, it follows
that 
\begin{align}
 & \left(\ln R_{1}\right)'(r_{\alpha})=\nonumber \\
 & =\frac{\left|\Delta_{\ell}\right|}{r_{\alpha}}+\frac{2\gamma_{\alpha}}{r_{\alpha}}\frac{M'\left(\frac{1}{2}\left|\Delta_{\ell}\right|+\frac{q^{2}}{2\alpha\gamma}+\frac{1}{2},\ell+1;\gamma_{\alpha}\right)}{M\left(\frac{1}{2}\left|\Delta_{\ell}\right|+\frac{q^{2}}{2\alpha\gamma}+\frac{1}{2},\ell+1;\gamma_{\alpha}\right)}>0\label{eq:log-deriv-lhs-whole}
\end{align}
for $q>0$. We conclude that Eq.~\eqref{eq:cond-reg1} is not satisfied
and that Eq.~\eqref{eq:log-deriv} has no solutions for $q>0$, implying
that there are no bound states of angular momenta $\ell\geq\gamma_{\alpha}$.

\subsection{Onset of bound states in finite size\label{subsec:Onset-of-bound}}

\subsubsection{Case $p=0$}

Although~\eqref{eq:eff-bound-st-criter} has, in general, no closed
form solution, we can obtain an approximate solution for the state
$p=0$ of each angular momentum $\ell$, since the onset of that state
will take place for small values of $\lambda_{\ell}$ whenever $R\gg\xi$.
We may thus solve
\begin{equation}
\Lambda_{\ell*}=\frac{\xi}{\alpha R}\label{eq:finite-size-onset}
\end{equation}
to leading order in $\lambda_{\ell}$, with $\Lambda_{\ell*}\equiv\Lambda_{\ell}(\lambda_{\ell*})$
the function given by Eq.~\eqref{eq:CapLambda_ell}

The function $\theta_{\ell}$ specified in Eq.~\eqref{eq:CapTheta}
can be Taylor-expanded around $\lambda_{\ell}\geq0$ to leading order
as
\[
\theta_{\ell}=\begin{cases}
\frac{\pi}{2}-\gamma_{\alpha}+\mathcal{O}(\gamma_{\alpha}^{2}) & ,\,\ell=0\\
-a_{\ell}\lambda_{\ell}+\mathcal{O}(\lambda_{\ell}^{3}) & ,\,\ell\geq1
\end{cases}\,,
\]
with
\[
a_{\ell}=\left(1+\frac{1}{\ell}\right)\frac{M\left(\frac{1}{2},\ell+1;\ell\right)}{M\left(\frac{3}{2},\ell+2;\ell\right)}\,,
\]
by virtue of the derivative of the CHF satisfying~\cite{Olver:2010:NHM:1830479}
\[
M'(a,b;z)=\frac{a}{b}M(a+1,b+1;z)\,.
\]
Note that, for $\ell\geq1$, we must first express $\gamma_{\alpha}$
to leading order in $\lambda_{\ell}$, given by $\gamma_{\alpha}=\ell+\lambda_{\ell}^{2}/(2\ell)+\mathcal{O}(\lambda_{\ell}^{4})$.
In turn, Eq.~\eqref{eq:varphi_leading} gives $\varphi_{\ell}=-\gamma_{\text{E}}\lambda_{\ell}+\mathcal{O}(\lambda_{\ell}^{3})$
to leading order in $\lambda_{\ell}$.

We begin with the case $\ell\geq1$. Eq.~\eqref{eq:finite-size-onset}
becomes, to leading order in $\lambda_{\ell}$,
\[
\frac{\sqrt{2}}{\ell}\exp\left(-\frac{\pi}{\lambda_{\ell*}}\right)e^{a_{\ell}-\gamma_{E}}=\frac{\xi}{\alpha R}\,,
\]
which can be rearranged to yield
\[
\lambda_{\ell*}=\frac{\pi}{\log\left(\frac{R}{\xi r_{0,\ell}}\right)}\left(1+\mathcal{O}(\lambda_{\ell*}^{2})\right)\,,
\]
with $r_{p=0,\ell}=e^{\gamma_{\text{E}}-a_{\ell}}\sqrt{3\ell^{2}/5}$.
Squaring this result, Eq.~\eqref{eq:eff-bound-st-criter} yields
\begin{equation}
\frac{g_{12}m_{2}}{g_{11}m_{1}}>\alpha^{2}\ell^{2}+\frac{c_{0,\ell}^{-2}}{\log\left(\frac{R}{\xi r_{0,\ell}}\right)^{2}}\,,\label{eq:onset_ellgeq1}
\end{equation}
with $c_{p=0,\ell}=\pi^{-1}\sqrt{6/5}$. Higher-order terms in $\log\left(R/r_{0,\ell}\right)^{-1}$
can be obtained by expanding $\theta_{\ell}$, $\varphi_{\ell}$ and
$\gamma_{\alpha}$ to subleading order in $\lambda_{\ell}$, but these
are found to be negligible.
\begin{table}[t]
\begin{centering}
\caption{Results of the fits~\eqref{eq:onset-fit} to the data from~\eqref{eq:onset-data}
for values of $c_{p,\ell}$ (top) and $\ln(1/r_{p,\ell})$ (bottom).\label{tab:Results-of-the}}
\par\end{centering}
\centering{}%
\begin{tabular*}{0.95\columnwidth}{@{\extracolsep{\fill}}@{\extracolsep{\fill}}@{\extracolsep{\fill}}cccccc}
\hline \hline
\noalign{\vskip0.05cm}
$p$ & $0$  & $1$ & $2$ & $3$ & $4$\tabularnewline
\hline 
\noalign{\vskip0.05cm}
\multirow{2}{*}{$\:\ell=0$} & 0.7251  & 0.2365  & 0.1357 & 0.0932 & 0.0706\tabularnewline
 & 1.0826 & 0.2642 & 0.2493 & 0.4039 & 0.5555\tabularnewline
\noalign{\vskip0.1cm}
\multirow{2}{*}{$\:\ell=1$} & 0.3248  & 0.1674 & 0.1130 & 0.0843 & 0.0663\tabularnewline
 & 1.0191 & 0.6487 & 0.4144 & 0.3376 & 0.3721\tabularnewline
\noalign{\vskip0.1cm}
\multirow{2}{*}{$\:\ell=2$} & 0.3133 & 0.1554 & 0.1042 & 0.0786 & 0.0629\tabularnewline
 & 0.1441 & 0.2474 & 0.2653 & 0.2562 & 0.2713\tabularnewline
\hline \hline
\noalign{\vskip0.1cm}
\end{tabular*}
\end{table}
\begin{figure}[t]
\includegraphics{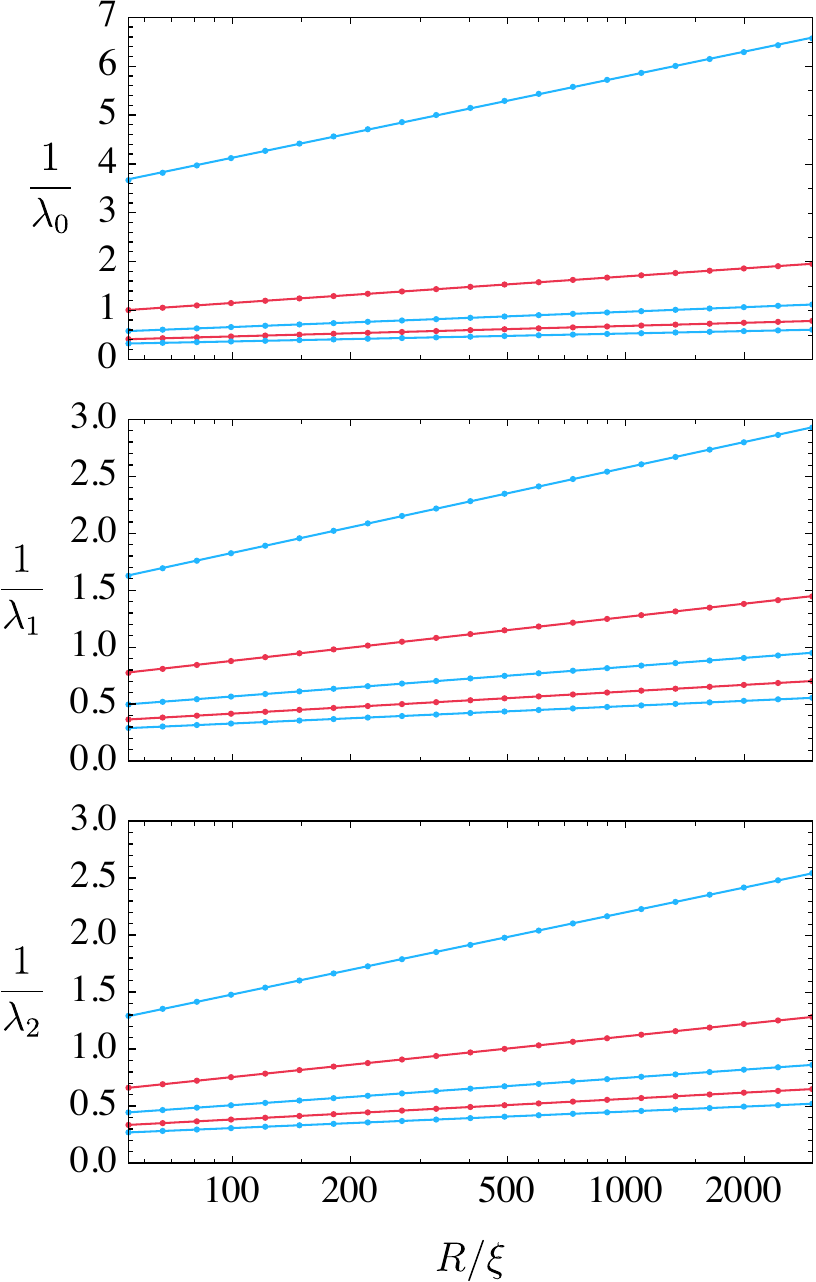}\caption{(color online) Comparisons of the fits~\eqref{eq:onset-fit} to the
data from~\eqref{eq:onset-data} for $\ell=0$ (top), $\ell=1$ (middle)
and $\ell=2$ (bottom), for $p=0,1,2,3,4$ in descending order on the
plots for each $\ell$. The horizontal axis $R/\xi$ is in logarithmic
scale.\label{fig:(color-online)-Comparisons}}
\end{figure}

For the case $\ell=0$, we may take Eq.~\eqref{eq:finite-size-onset}
to leading order in $\gamma_{\alpha}$ and rearrange it as
\begin{equation}
\gamma_{\alpha*}=\frac{3\pi}{2}\frac{1}{1-\gamma_{E}-\log\frac{\xi\gamma_{\alpha*}}{\sqrt{2}\alpha R}}\,.\label{eq:sat-eq_elleq0}
\end{equation}
We note that for $R\gg\xi$, the RHS of~\eqref{eq:sat-electable}
depends very weakly on $\gamma_{\alpha*}$; indeed, numerical evidence
indicates that the solution $\gamma_{\alpha*}=\gamma_{\alpha*}(R)$,
for each $R$, is well approximated to zeroth order in $R$ by $\gamma_{\alpha*}(R)\approx\gamma_{\text{E}}$
over a range $200<R/\xi<2000$~. Then, we may write $\gamma_{\alpha*}=\gamma_{\text{E}}+\gamma_{\text{E}}\delta_{*}$,
with $\delta_{*}=\gamma_{\alpha}/\gamma_{\text{E}}-1\ll1$ small,
by hypothesis, so that we can take~\eqref{eq:sat-eq_elleq0} to leading
order in $\delta_{*}$ to yield 
\begin{equation}
\gamma_{\alpha*}=\frac{\frac{3\pi}{2}-\gamma_{\text{E}}}{\log\left(\frac{R}{\xi r_{0,0}}\right)}\,,
\end{equation}
with $r_{0,0}=e^{\gamma_{\text{E}}}\sqrt{3\gamma_{\text{E}}^{2}/5}$.
Finally, squaring this result, Eq.~\eqref{eq:eff-bound-st-criter}
yields
\begin{equation}
\frac{g_{12}m_{2}}{g_{11}m_{1}}>\frac{c_{0,0}^{-2}}{\log\left(\frac{R}{\xi r_{0,0}}\right)^{2}}\,,\label{eq:onset-elleq0}
\end{equation}
with $c_{0,0}=\left(3\pi/2-\gamma_{\text{E}}\right)^{-1}\sqrt{6/5}$.
These results suggest we can establish a law
\begin{equation}
\frac{g_{12}m_{2}}{g_{11}m_{1}}>\alpha^{2}\ell^{2}+\frac{c_{0,\ell}^{2}}{\log\left(\frac{R}{\xi r_{0,\ell}}\right)^{2}}\,,\label{eq:onset-peq0}
\end{equation}
valid, at least, for $p=0$.

\subsubsection{Case $p\geq0$}

In general, the foregoing considerations in the derivations of~\eqref{eq:onset_ellgeq1}
and~\eqref{eq:onset-elleq0} no longer hold. However, we can show
that condition~\eqref{eq:onset-peq0} is accurate as well for $p\geq1$:
we obtain numerical solutions of Eq.~\eqref{eq:finite-size-onset}
over a range $50<R/\xi<3000$, by employing a Newton method to the
equation
\begin{equation}
\ln\Lambda_{\ell*}=-\ln\frac{\alpha R}{\xi}\,;\label{eq:onset-data}
\end{equation}
we then perform a linear fit of the expression on the RHS of~\eqref{eq:onset-peq0}
to parameters $c_{p,\ell}$ and $r_{p,\ell}$ in a $1/\lambda_{\ell}$
vs. $\ln(R/\xi)$ plot, i.e.
\begin{equation}
\frac{1}{\lambda_{\ell}}=c_{p,\ell}\left(\ln\frac{R}{\xi}-\ln r_{p,\ell}\right)\,.\label{eq:onset-fit}
\end{equation}
Graphical comparisons are in displayed in Fig.~\ref{fig:(color-online)-Comparisons},
while results of the fit are given in Table.~\ref{tab:Results-of-the}
for a few bound states. We thus conclude that Eq.~\eqref{eq:p0-condition-boundst}
provides a condition for the onset of any bound state $(p,\ell)$.

\bibliographystyle{apsrev4-1}
\bibliography{vortex_imp.bib}

\begin{thebibliography}{67}
\bibitem [{foo({\natexlab{a}})}]{footnote_1}%
  \BibitemOpen
  \href@noop {} {\bibfield  {journal} {\bibinfo  {journal} {In the present
  work, we avoid the usual notation for the incomplete gamma function as to
  avoid any ambiguities with the parameter $\gamma$.}\ }
  ({\natexlab{a}})}\BibitemShut {NoStop}%
\bibitem [{foo({\natexlab{b}})}]{footnote_2}%
  \BibitemOpen
  \href@noop {} {\bibfield  {journal} {\bibinfo  {journal} {We disregard the
  condition $\text{Re}\beta>0$ by the convergence argument presented following
  Eq.~\eqref{eq:stationary-g} and due to the fact that the integral in
  Eq.~\eqref{eq:integration-saddle-pt-1} exists formally even for
  $\text{Re}\beta=0$.}\ } ({\natexlab{b}})}\BibitemShut {NoStop}%
\end{thebibliography}
\bibliographystyle{apsrev4-1}

\end{document}